\documentclass{ws-mpla}

\def\be{\begin{equation}}
\def\ee{\end{equation}}
\def\ba{\begin{eqnarray}}
\def\ea{\end{eqnarray}}

\def\ga{\mathrel{\raise.3ex\hbox{$>$\kern-.75em\lower1ex\hbox{$\sim$}}}}
\def\la{\mathrel{\raise.3ex\hbox{$<$\kern-.75em\lower1ex\hbox{$\sim$}}}}
\def\apj #1 #2 #3 {#1, ApJ, {\bf #2}, #3}
\def\apjl #1 #2 #3 {#1, ApJ, {\bf #2}, L#3}
\def\apjs #1 #2 #3 {#1, ApJS, {\bf #2}, #3}
\def\aap  #1 #2 #3 {#1, A\&A, {\bf #2}, #3}
\def\mnras #1 #2 #3 {#1, MNRAS, {\bf #2}, #3}
\def\pra #1 #2 #3 {#1, Phys.~Rev.~A., {\bf #2}, #3}
\def\prb #1 #2 #3 {#1, Phys.~Rev.~B., {\bf #2}, #3}
\def\prc #1 #2 #3 {#1, Phys.~Rev.~C., {\bf #2}, #3}
\def\prd #1 #2 #3 {#1, Phys.~Rev.~D., {\bf #2}, #3}
\def\pre #1 #2 #3 {#1, Phys.~Rev.~E., {\bf #2}, #3}
\def\prl #1 #2 #3 {#1, Phys.~Rev.~Lett., {\bf #2}, #3}
\def\plb #1 #2 #3 {#1, Phys.~Lett.~B., {\bf #2}, #3}
\def\science #1 #2 #3 {#1, Science., {\bf #2}, #3}
\def\nature #1 #2 #3 {#1, Nature., {\bf #2}, #3}
\def\nphysa #1 #2 #3 {#1, Nucl.~Phys.~A., {\bf #2}, #3}
\def\nphysb #1 #2 #3 {#1, Nucl.~Phys.~B., {\bf #2}, #3}
\def\nphysbs #1 #2 #3 {#1, Nucl.~Phys.~B.~Suppl., {\bf #2}, #3}

\def\h#1{\hbox{${}^{#1}$H}}

\def\be#1{\hbox{${}^{#1}$Be}}

\def\h502{\hbox{$ h^{2}_{50}$}}

\def\la{\mathrel{\mathpalette\fun <}}
\def\ga{\mathrel{\mathpalette\fun >}}
\def\fun#1#2{\lower3.6pt\vbox{\baselineskip0pt\lineskip.9pt
  \ialign{$\mathsurround=0pt#1\hfil##\hfil$\crcr#2\crcr\sim\crcr}}}

\newcommand{\vct}[1]{\mbox{\boldmath${#1}$}}

\begin{document}

\markboth{Yong-Yeon Keum} {Probing for dynamics of Dark Energy in
Mass-Varying-Neutrinos(MaVaNu): CMB and LSS }

\catchline{}{}{}{}{}

\title{Probing for Dynamics of Dark-Energy in Mass Varying Neutrinos:
\\ Cosmic Microwave Background Radiation and Large Scale Structure}
\author{\footnotesize Yong-Yeon Keum}

\address{Department of Physics, National Taiwan University, \\ No.1 Sec.4, Roosevelt Road, Taipei, Taiwan, 10672, R.O.C.\\
yykeum@phys.ntu.edu.tw}

\maketitle

\pub{Received (30 January 2007)}{Revised (25 April 2007)}

\begin{abstract}
We present cosmological perturbation theory in neutrino probe
interacting dark-energy models, and calculate cosmic microwave
background anisotropies and matter power spectrum. In these models,
the evolution of the mass of neutrinos is determined by the
quintessence scalar field, which is responsible for the cosmic
acceleration today. We consider several types of scalar field
potentials and put constraints on the coupling parameter between
neutrinos and dark energy. Assuming the flatness of the universe,
the constraint we can derive from the current observation is $\sum
m_{\nu} < 0.87 eV$ at the 95 $\%$ confidence level for the sum over
three species of neutrinos.

\end{abstract}

\ccode{PACS Nos.: 98.80.-k,98.80.Jk,98.80.Cq}

\section{Introduction}
After SNIa\cite{sn1a} and WMAP\cite{wmap} observations during last
decade, the discovery of the accelerated expansion of the universe
is a major challenge of particle physics and cosmology. There are
currently three candidates for the {\it Dark-Energy} which derives
this accelerated expansion:
\begin{itemize}
\item a non-zero cosmological constant\cite{lambda},
\item a dynamical cosmological constant (Quintessence scalar field)\cite{quintessence},
\item modifications of Einstein Theory of Gravity\cite{mgrav}
\end{itemize}

In this paper, we review shortly the main idea of three possible
candidates and their cosmological phenomena. Specially we consider
the interacting mechanism between dark-energy with a hot dark-matter
(neutrinos). Within neutrinos probe interacting dark-energy
scenario\cite{mavanu}, we calculate Cosmic Microwave Background(CMB)
radiation and Large Scale Structure(LSS) within cosmological
perturbation theory. The evolution of the mass of neutrinos is
determined by the quintessence scalar filed, which is responsible
for the cosmic acceleration today.

\section{Three possible solutions for Accelerating Universe:}
Recent observations with Supernova Ia type (SNIa) and CMB radiation
have provided strong evidence that we live now in an accelerating
and almost flat universe. In general, one believes that the
dominance of a dark-energy component with negative pressure in the
present era is responsible for the universe's accelerated expansion.
However there are three possible solutions to explain the
accelerating universe. The Einstein Equation in General Relativity
is given by the following form:
\begin{equation}
G_{\mu\nu} \,\,= \,\, R_{\mu\nu} -{1 \over 2} R \, g_{\mu\nu} \hspace{2mm}
= \hspace{2mm} 8 \pi G \, T_{\mu\nu} \hspace{2mm} + \hspace{2mm} \Lambda \, g_{\mu\nu},
\label{eq:einstein}
\end{equation}
Here, $G_{\mu\nu}$ term contains the information of geometrical
structure, the energy-momentum tensor $T_{\mu\nu}$ keeps the
information of matter distributions, and the last term is so called
the cosmological constant which contain the information of non-zero
vacuum energy. After solve the Einstein equation, one can drive a
simple relation:
\begin{equation}
{\ddot{R} \over R} = -{4 \pi G \over 3} (\rho +3p) +{\Lambda \over
3}.
\end{equation}
In order to get the accelerating expansion, either cosmological
constant $\Lambda$ ($\omega_{\Lambda} = P/\rho=-1$) becomes positive
or a new concept of dark-energy with the negative pressure
($\omega_{\phi} <-1/3$) needs to be introduced. Another solution can
be given by the modification of geometrical structure which can
provide a repulsive source of gravitational force.  In this case,
the attractive gravitational force term is dominant in early stage
of universe, however at later time near the present era, repulsive
term become important and drives universe to be expanded with an
acceleration.
 Also we can consider extra-energy density
contributions from bulk space in Brane-World scenario models, which
can modify the Friedmann equation as $H^2 \propto \rho + \rho^{'}$.
In summary, we have three different solutions for the accelerating
expansion of our universe as mentioned in the introduction.
 Probing for the origin of accelerating universe is the most
important and challenged problem in high energy physics and
cosmology now. The detail explanation and many references are in a
useful review on dark energy\cite{review-DE}.

In this paper, we concentrate on the second solution using the
quintessence field. In present epoch, the potential term becomes
important than kinetic term, which can easily explain the negative
pressure with $\omega_{\phi}^0 \simeq -1$. However there are many
different versions of quintessence field: K-essence\cite{k-essence},
phantom\cite{phantom}, quintom\cite{quintom}, ....etc., and to
justify the origin of dark-energy from experimental observations is
really a difficult job. Present updated value of the equation of
states(EoS) are $\omega=-1.02 \pm 0.12$ without any supernova
data\cite{seljak:0604335}.

\section{Interacting Dark-Energy with Neutrinos:}
As explained in previous section, it is really difficult to probe
the origin of dark-energy when the dark-energy doesn't interact with
other matters at all. Here we investigate the cosmological
implication of an idea of the dark-energy interacting with neutrinos
\cite{{mavanu},{Fardon:2003eh}}. For simplicity, we consider the
case that dark-energy and neutrinos are coupled such that the mass
of the neutrinos is a function of the scalar field which drives the
late time accelerated expansion of the universe. In previous works
by Fardon et al.\cite{Fardon:2003eh} and R. Peccei\cite{mavanu},
kinetic energy term was ignored and potential term was treated as a
dynamical cosmology constant, which can be applicable for the
dynamics near present epoch. However the kinetic contributions
become important to descreibe cosmological perturbations in early
stage of universe, which is fully considered in our analysis.

\subsection{Cosmological perturbations}
Equations for quintessence scalar field are given by
\begin{eqnarray}
\ddot{\phi}&+&2{\cal H}\dot\phi+a^2\frac{d V_{\rm eff}(\phi)}{d\phi}=0~,
 \label{eq:Qddot}\\
V_{\rm eff}(\phi)&=&V(\phi)+V_{\rm I}(\phi)~,\\
V_{\rm I}(\phi)&=&a^{-4}\int\frac{d^3q}{(2\pi)^3}\sqrt{q^2+a^2
 m_\nu^2(\phi)}f(q)~,\\
m_\nu(\phi) &=& \bar m_i e^{\beta\frac{\phi}{M_{\rm pl}}}~{\mbox{(as an example)}},
\end{eqnarray}
where $V(\phi)$ is the potential of quintessence scalar field,
$V_{\rm I}(\phi)$ is additional potential due to the coupling to
neutrino particles \cite{Fardon:2003eh,Bi:2003yr}, and $m_\nu(\phi)$
is the mass of neutrino coupled to the scalar field. ${\cal H}$ is
${\dot a}/a$, where the dot represents the derivative with respect
to the conformal time $\tau$.

Energy densities of mass varying neutrino (MVN) and quintessence scalar
field are described as
\begin{eqnarray}
\rho_\nu &=& a^{-4}\int \frac{d^3 q}{(2\pi)^3} \sqrt{q^2+a^2m_\nu^2} f_0(q)~, \label{eq:rho_nu}\\
3P_\nu &=& a^{-4}\int \frac{d^3 q}{(2\pi)^3} \frac{q^2}{\sqrt{q^2+a^2m_\nu^2}}
 f_0(q)~, \label{eq:P_nu}\\
\rho_\phi &=& \frac{1}{2a^2}\dot\phi^2+V(\phi)~,\\
P_\phi &=& \frac{1}{2a^2}\dot\phi^2-V(\phi)~.
\end{eqnarray}
From equations (\ref{eq:rho_nu}) and (\ref{eq:P_nu}), the equation of
motion for the background energy density of neutrinos is given by
\begin{equation}
\dot\rho_{\nu}+3{\cal H}(\rho_\nu+P_\nu)=\frac{\partial \ln
 m_\nu}{\partial \phi}\dot\phi(\rho_\nu -3P_\nu)~.
\end{equation}
In our analysis, we are working in the synchronous gauge with line element:
\begin{equation}
 ds^2 = a^2(\tau)\left[-d\tau^2 + (\delta_{ij}+h_{ij})dx^i dx^j\right]~,
\end{equation}
For CMB anisotropies we mainly consider the scalar type
perturbations. We introduce two scalar fields, $h(\vct k, \tau)$ and
$\eta(\vct k, \tau)$, in k-space and write the scalar mode of
$h_{ij}$ as a Fourier integral \cite{Ma:1995ey}
\begin{equation}
h_{ij}(\vct x,\tau)=\int d^3k e^{i\vct{k}\cdot\vct{x}}\left[\vct{\hat
 k}_i\vct{\hat k}_j h(\vct k,\tau)+(\vct{\hat k}_i \vct{\hat k}_j
 -\frac{1}{3}\delta_{ij})6\eta(\vct{k},\tau) \right]~,
\end{equation}
where $\vct{k} =k\vct{\hat k}$ with $\hat{k}^i \hat{k}_i =1$.

The equation of quintessence scalar field is given by
\begin{equation}
 \Box \phi - V_{\rm eff}(\phi)=0~.
\end{equation}
Let us write the scalar field as a sum of background value and
perturbations around it, $\phi(\vct x,\tau)=\phi(\tau)+\delta \phi(\vct
x,\tau)$.
The perturbation equation is then described as
\begin{equation}
\frac{1}{a^2}\ddot{\delta \phi}+\frac{2}{a^2}{\cal H}\dot\delta\phi -
 \frac{1}{a^2}\nabla^2(\delta\phi)+\frac{1}{2a^2}\dot{h}\dot\phi+\frac{d^2
 V}{d\phi^2}\delta\phi +\delta\left(\frac{dV_{\rm I}}{d\phi}\right)=0~ \label{eq:dQddot},
\end{equation}
To describe $\delta\left(\frac{dV_{\rm I}}{d\phi}\right)$,
we shall write the distribution function of neutrinos with background
distribution and perturbation around it as
\begin{equation}
f(x^i, \tau, q, n_j )= f_0(\tau,q)(1+\Psi(x^i, \tau, q, n_j))~.
\end{equation}

After some calculations, we finally obtain the useful equations\cite{kiyo-keum}:
\begin{eqnarray}
\frac{dV_{\rm I}}{d\phi} &=& \frac{\partial \ln m_\nu}{\partial
 \phi}(\rho_\nu -3P_\nu)~, \\
\delta\left(\frac{dV_{\rm I}}{d\phi}\right) &=&
\frac{\partial^2 \ln m_\nu}{\partial
 \phi^2} \delta\phi (\rho_\nu-3P_\nu) \nonumber \\
&&+ \frac{\partial \ln m_\nu}{\partial
 \phi}(\delta\rho_\nu-3\delta P_\nu)
\end{eqnarray}
Note that perturbation fluid variables in mass varying neutrinos are given by
\begin{eqnarray}
\delta\rho_\nu&=&a^{-4}\int\frac{d^3 q}{(2\pi)^3} \epsilon
 f_0(q)\Psi+a^{-4}\int\frac{d^3 q}{(2\pi)^3}\frac{\partial
 \epsilon}{\partial \phi}\delta\phi f_0 ~, \label{eq:delta_rho_nu}\\
3\delta P_\nu&=&a^{-4}\int\frac{d^3 q}{(2\pi)^3} \frac{q^2}{\epsilon}
 f_0(q)\Psi-a^{-4}\int\frac{d^3
 q}{(2\pi)^3}\frac{q^2}{\epsilon^2}\frac{\partial \epsilon}{\partial
 \phi}\delta\phi f_0 ~.
\end{eqnarray}

\subsection{Boltzmann Equation}
The Boltzmann equation is given in general,
\begin{equation}
 \frac{Df}{D\tau}=\frac{\partial f}{\partial
 \tau}+ \frac{dx^i}{d\tau}\frac{\partial f}{\partial
 x^i}
+\frac{dq}{d\tau}\frac{\partial f}{\partial
 q}+\frac{dn_i}{d\tau}\frac{\partial f}{\partial
 n_i}=\left(\frac{\partial f}{\partial \tau}\right)_C~.
\end{equation}
From the time component of geodesic equation \cite{Anderson:1997un},
\begin{equation}
 \frac{1}{2}\frac{d}{d\tau}\left(P^0\right)^2=-\Gamma^0_{\alpha\beta}P^\alpha
  P^\beta - m g^{0 \nu}m_{,\nu}~,
\label{eq:eq-a}
\end{equation}
and the relation $P^0=a^{-2}\epsilon=a^{-2}\sqrt{q^2+a^2 m_\nu^2}$, we
have
\begin{equation}
 \frac{dq}{d\tau}=-\frac{1}{2}\dot{h_{ij}}qn^in^j-a^2
  \frac{m}{q}\frac{\partial m}{\partial x^i}\frac{dx^i}{d \tau} ~.
\label{eq:eq-b}
\end{equation}
Our analytic formulas in eqs.(\ref{eq:eq-a}-\ref{eq:eq-b})
 are completely different from those of Brookfield et al.\cite{Brookfield-b},
since they have missed the contribution of the varying neutrino mass
term. In later this term also give an important contribute in the
first order perturbation of the Boltzman equation. The detail
calculations will be shown in elsewhere\cite{kiyo-keum}.

The zeroth-order Boltzmann equation is given by
\begin{equation}
 \frac{\partial f_0}{\partial\tau}=0~.
\label{eq:Boltz0}
\end{equation}
The Fermi-Dirac distribution
\begin{equation}
 f_0=f_0(\epsilon)=\frac{g_s}{h_{\rm P}^3}\frac{1}{e^{\epsilon/k_{\rm B}T_0}+1}~,
\end{equation}
can be a solution.
Here $g_s$ is the number of spin degrees of freedom, $h_{\rm P}$ and
$k_{\rm B}$ are the Planck and the Boltzmann constants. We assume that
MVNs are decoupled from the thermal bath when they are extremely
relativistic so we can simply replace $\epsilon$ in the unperturbed
Fermi-Dirac distribution by $q$. Thus we have
\begin{equation}
 f_0=f_0(\epsilon)=\frac{g_s}{h_{\rm P}^3}\frac{1}{e^{q/k_{\rm B}T_0}+1}~,
\end{equation}
whish can also be a solution of eq.(\ref{eq:Boltz0}).

The first-order Boltzmann equation is
\begin{eqnarray}
 \frac{\partial \Psi}{\partial
 \tau}&+&i\frac{q}{\epsilon}(\vct{\hat{n}}\cdot\vct{k})\Psi+\left(\dot\eta-(\vct{\hat
 k}\cdot\vct{\hat n})^2\frac{\dot h+6\dot\eta}{2}\right)\frac{\partial \ln
 f_0}{\partial \ln q}\nonumber \\
&-&i\frac{q}{\epsilon}(\vct{\hat{n}}\cdot\vct{k})k\delta\phi\frac{a^2
 m^2}{q^2}\frac{\partial \ln m}{\partial \phi}\frac{\partial \ln
 f_0}{\partial \ln q} = 0~.
\label{eq:boltzmann}
\end{eqnarray}
Following previous studies, we shall assume that the initial momentum
dependence is axially symmetric so that $\Psi$ depends on
$\vct{q}=q\vct{\hat n}$ only through $q$ and $\vct{\hat k}\cdot\vct{\hat
n}$.
With this assumption, we expand the perturbation of distribution
function, $\Psi$, in a Legendre series,
\begin{equation}
 \Psi(\vct{k},\vct{\hat n},q,\tau)=\sum (-i)^\ell(2\ell+1)\Psi_\ell(\vct{k},q,\tau)P_\ell(\vct{\hat{k}}\cdot\vct{\hat{n}})~.
\end{equation}
Then we obtain the hierarchy for MVN
\begin{eqnarray}
 \dot{\Psi_0}&=&-\frac{q}{\epsilon}k\Psi_1
                +\frac{\dot h}{6}\frac{\partial \ln{f_0}}{\partial\ln{q}}~, \label{eq:dot_Psi_0}\\
\dot{\Psi_1}&=&\frac{1}{3}\frac{q}{\epsilon}k\left(\Psi_0-2\Psi_2\right)
              + \kappa~, \label{eq:dot_Psi_1}\\
\dot{\Psi_2}&=&\frac{1}{5}\frac{q}{\epsilon}k(2\Psi_1-3\Psi_3)-\left(\frac{1}{15}\dot{h}+\frac{2}{5}\dot{\eta}\right)\frac{\partial \ln{f_0}}{\partial \ln{q}}~,\\
\dot{\Psi_\ell}&=&\frac{q}{\epsilon}k\left(\frac{\ell}{2\ell+1}\Psi_{\ell-1}-\frac{\ell+1}{2\ell+1}\Psi_{\ell+1}\right)~.
\end{eqnarray}
where
\begin{equation}
\kappa = -\frac{1}{3}\frac{q}{\epsilon}k\frac{a^2 m^2}{q^2}\delta\phi
 \frac{\partial \ln m_\nu}{\partial \phi}\frac{\partial \ln
 f_0}{\partial \ln q}\label{eq:kappa}~.
\end{equation}
Here we used the recursion relation
\begin{equation}
 (\ell+1)P_{\ell+1}(\mu)=(2\ell+1)\mu P_\ell(\mu)-\ell P_{\ell-1}(\mu)~.
\end{equation}
We have to solve these equations with a $q$-grid for every wavenumber $k$.

\subsection{Quintessence potentials}
To determine the evolution of scalar field which couples to
neutrinos, we should specify the potential of the scalar field. A
variety of quintessence effective potentials can be found in the
literature. In this paper we examine three type of quintessential
potentials. First we analyze what is a frequently invoked form for
the effective potential of the tracker field, i.e., an inverse power
law such as originally analyzed by Ratra and Peebles
\cite{Ratra:1987rm},
\begin{equation}
V(\phi)=M^{4+\alpha}\phi^{-\alpha}~~~\mbox{(Model I)}~,
\end{equation}
where $M$ and $\alpha$ are parameters.

We will also consider a modified form of $V(\phi)$ as proposed by
Brax and Martin \cite{Brax:1999gp} based on the condition that the
quintessence fields be part of supergravity models. The potential
now becomes
\begin{equation}
V(\phi)=M^{4+\alpha}\phi^{-\alpha} e^{3\phi^2/2m_{\rm
pl}^2}~~~\mbox{(Model II)}~,
\end{equation}
where the exponential correction becomes important near the present
time as $\phi \to m_{\rm pl}$. The fact that this potential has a
minimum for $\phi=\sqrt{\alpha/3}m_{\rm pl}$ changes the dynamics.
It causes the present value of $w$ to evolve to a cosmological
constant much quicker than for the bare power-law potential
\cite{Brax:2000yb}. In these models the parameter $M$ is fixed by
the condition that $\Omega_{\phi}\approx 0.7$ at present.

\begin{center}
\begin{figure}
\vspace{0cm}
\epsfxsize=5cm
\centerline{
    {\includegraphics[width=0.48\textwidth]{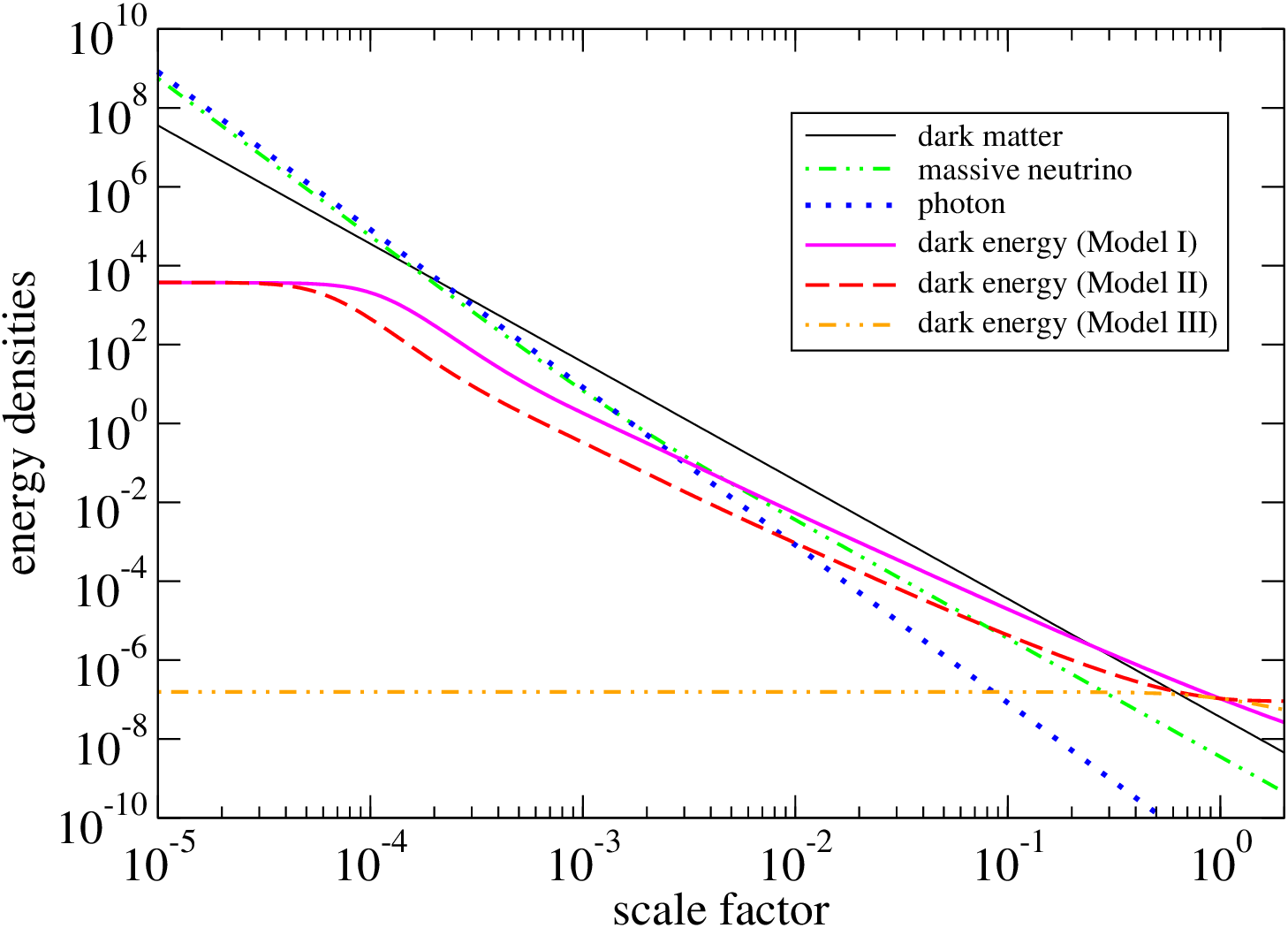}}
    {\includegraphics[width=0.48\textwidth]{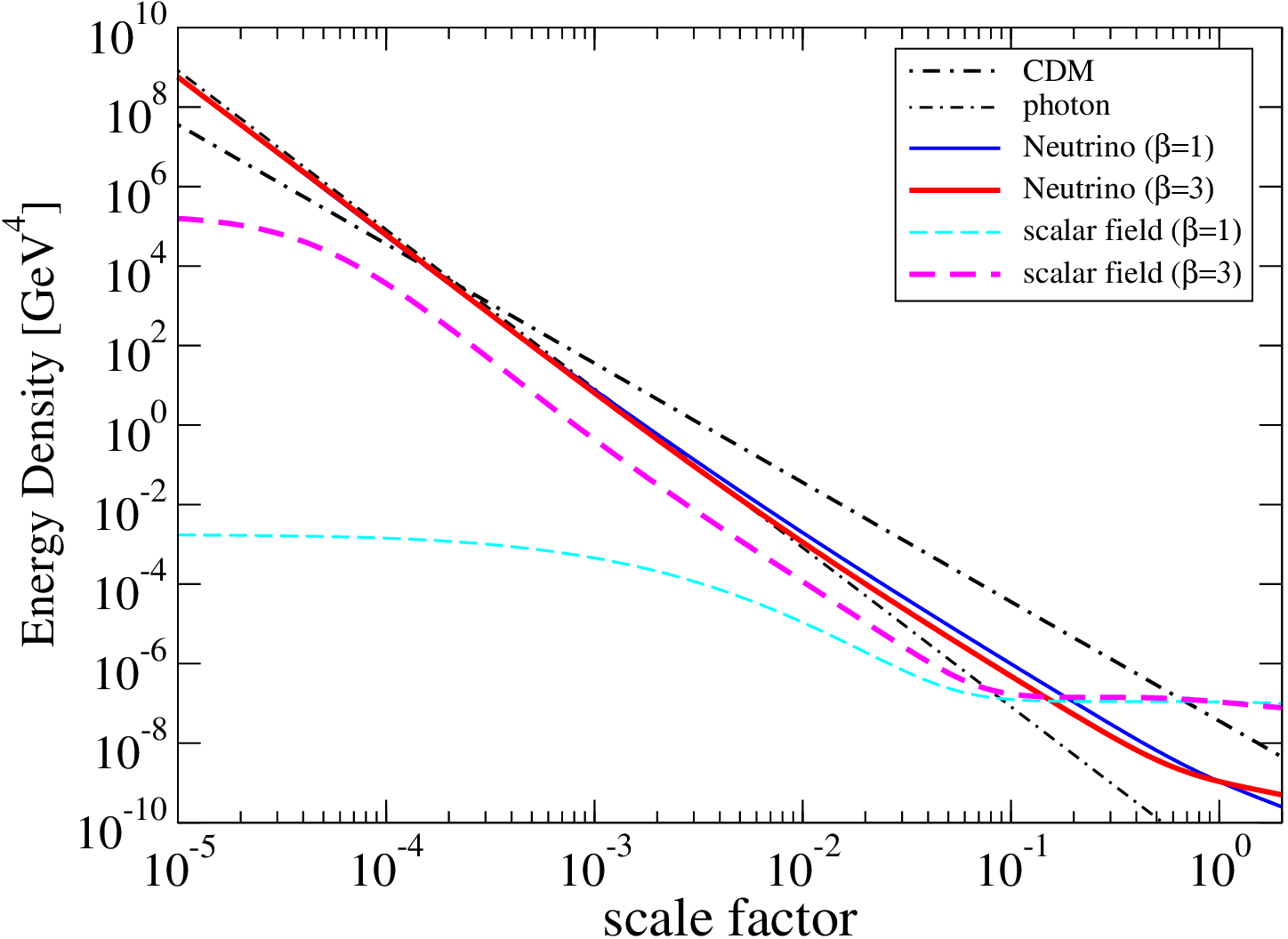}} }
\epsfxsize=5cm
\caption{Examples of the evolution of energy density in quintessence
and
 the background fields as indicated. Model parameters taken to plot this
 figure are $\alpha=10$, $10$, $1$ for model I, II, III,
 respectively. The other parameters for the dark energy are fixed so that
 the energy densities in three types of dark energy should be the same at
 present(left-handed side figure).} \label{fig:energy_densities}
\caption{Examples of the evolution of energy density in quintessence
and the background fields in coupled cases with inverse power law potential
 (Model I). Model parameters taken to plot this
 figure are $\alpha=1$, $\beta=1$, $3$ as indicated. The other
 parameters for the dark energy are fixed so that
 the energy densities in three types of dark energy should be the same at
 present(right-hand side figure).} \label{fig:energy_densities_coupled}
\end{figure}
\end{center}

We will also analyze another class of tracking potential, namely,
the potential of exponential type \cite{Copeland:1997et}:
\begin{equation}
V(\phi) = M^4 e^{-\alpha\phi}~~~\mbox{(Model III)}~,
\end{equation}
This type of potential can lead to accelerating expansion provided
that $\alpha<\sqrt{2}$. In figure (\ref{fig:energy_densities}), we
present examples of evolution of energy densities with these three
types of potentials with vanishing coupling strength to neutrinos.

\subsection{Time evolution of neutrino mass and energy density in scalar
 field}
\begin{figure}
\begin{center}
    \rotatebox{0}{\includegraphics[width=0.48\textwidth]{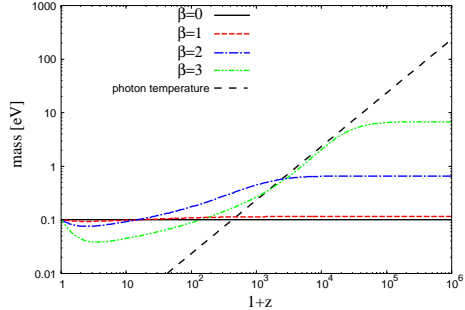}}
\end{center}
\caption{Examples of the time evolution of neutrino mass in power
law
 potential models (Model I) with $\alpha=1$ and $\beta=0$ (black solid line),
 $\beta=1$ (red dashed line), $\beta=2$ (blue dash-dotted line),
 $\beta=3$ (dash-dot-dotted line). The larger coupling parameter leads
 to the larger mass in the early universe.
} \label{fig:mass_evolution}
\end{figure}

For an illustration we also plot examples of evolution of energy
densities for interacting case with inverse power law potential
(Model I) in Fig. (\ref{fig:energy_densities_coupled}). In
interacting dark energy cases, the evolution of the scalar field is
determined both by its own potential and interacting term from
neutrinos. When neutrinos are highly relativistic, the interaction
term can be expressed as
\begin{equation}
\frac{\partial m_\nu}{\partial \phi}(\rho_\nu-3P_\nu)\approx
 \frac{10}{7\pi^2} (am_\nu)^2 \rho_{\nu_{\rm massless}}~,
\end{equation}
where $\rho_{\nu_{\rm massless}}$ denotes the energy density of
neutrinos with no mass. The term roughly scales as $\propto a^{-2}$,
and therefore, it dominates deep in the radiation dominated era.
However, because the motion of the scalar field driven by this
interaction term is almost suppressed by the friction term, $-3{\cal
H}\dot\phi$. The scalar field satisfies the slow roll condition
similar to the inflation models, $-3{\cal H}\dot\phi\approx
a^2\frac{\partial m_\nu}{\partial \phi}(\rho_\nu-3P_\nu)$. Thus, the
energy density in scalar field and the mass of neutrinos is frozen
there. These behaviors are clearly seen in Figs.
(\ref{fig:energy_densities_coupled}) and (\ref{fig:mass_evolution}).
\subsection{Constrains on the MaVaNu parameters}
As was shown in the previous sections, the coupling between
cosmological neutrinos and dark energy quintessence could modify the
CMB and matter power spectra significantly.  It is therefore
possible and also important to put constraints on coupling
parameters from current observations. For this purpose, we use the
WMAP3 \cite{Hinshaw:2006ia,Page:2006hz} and 2dF \cite{Cole:2005sx}
data sets.
\begin{center}
\begin{figure}
\vspace{0.3cm}
\epsfxsize=5cm
\centerline{\rotatebox{0}{\includegraphics[width=0.48\textwidth]{cl_spectra.eps}}
                          {\includegraphics[width=0.48\textwidth]{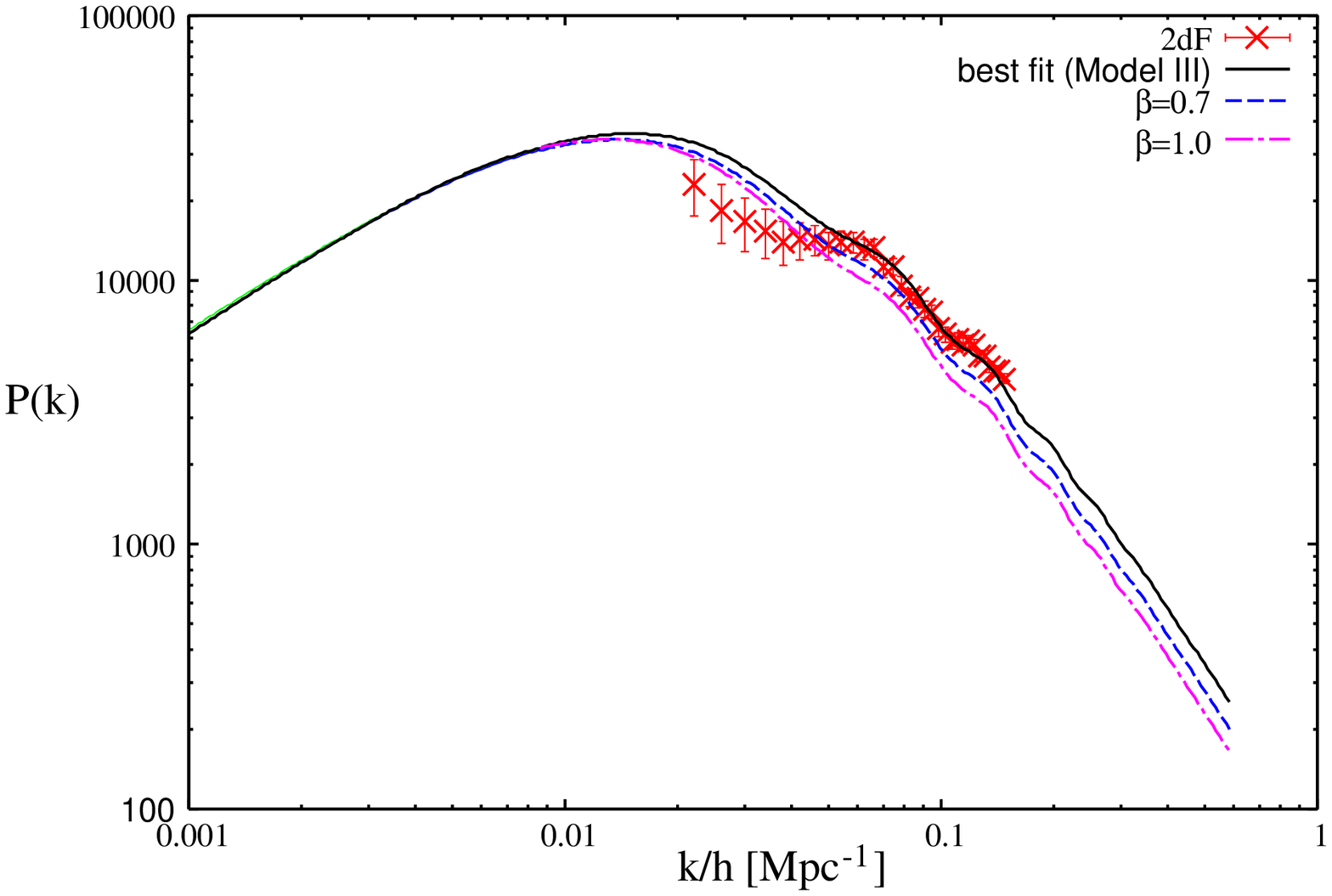}} }
\epsfxsize=5cm
\caption{(left-hand side figure)
The CMB angular power spectra for Model I. The solid line
is the best fit for the model ($(\alpha,\beta)=(2.97,0.170)$),
the other lines are models with different
 parameter value of $\alpha$ and $\beta$ as indicated.
The points are WMAP three year data. } \label{fig:cl_spectra}
\caption{(right-hand side figure)
The CMB angular power spectra for Model III. The solid line
is the best fit for the model ($(\alpha,\beta)=(0.78,0.28)$), the other
 lines are models with different
 parameter value of $\alpha$ and $\beta$ as indicated.
The points are 2dF data. } \label{fig:pk_spectra}
\end{figure}
\end{center}
\vspace{-0.3cm}

The flux power spectrum of the Lyman-$\alpha$ forest can be used to
measure the matter power spectrum at small scales around $z\la 3$
\cite{McDonald:1999dt,Croft:2000hs}. It has been shown, however,
that the resultant constraint on neutrino mass can vary
significantly from $\sum m_\nu < 0.2$eV to $0.4$eV depending on the
specific Lyman-$\alpha$ analysis used \cite{Goobar:2006xz}. The
complication arises because the result suffers from the systematic
uncertainty regarding to the model for the intergalactic physical
effects, i.e., damping wings, ionizing radiation fluctuations,
galactic winds, and so on \cite{McDonald:2004xp}. Therefore, we
conservatively omit the Lyman-$\alpha$ forest data from our
analysis.

Because there are many other cosmological parameters than the MaVaNu
parameters, we follow the Markov Chain Monte Carlo(MCMC) global fit
approach \cite{MCMC} to explore the likelihood space and marginalize
over the nuisance parameters to obtain the constraint on
parameter(s) we are interested in. Our parameter space consists of
\begin{equation}
\vec{P}\equiv
(\Omega_bh^2,\Omega_ch^2,H,\tau,A_s,n_s,m_i,\alpha,\beta)~,
\end{equation}
where $\omega_bh^2$ and $\Omega_ch^2$ are the baryon and CDM
densities in units of critical density, $H$ is the hubble parameter,
$\tau$ is the optical depth of Compton scattering to the last
scattering surface, $A_s$ and $n_s$ are the amplitude and spectral
index of primordial density fluctuations, and $(m_i,\alpha,\beta)$
are the parameters of MaVaNu defined in sections 3.1 and 3.3. We
have put priors on MaVaNu parameters as $\alpha>0$, and $\beta>0$
for simplicity and saving the computational time.
\begin{figure}
\vspace{0.0cm}
\begin{center}
\epsfxsize=5cm
\centerline{
    \rotatebox{0}{\includegraphics[width=0.48\textwidth]{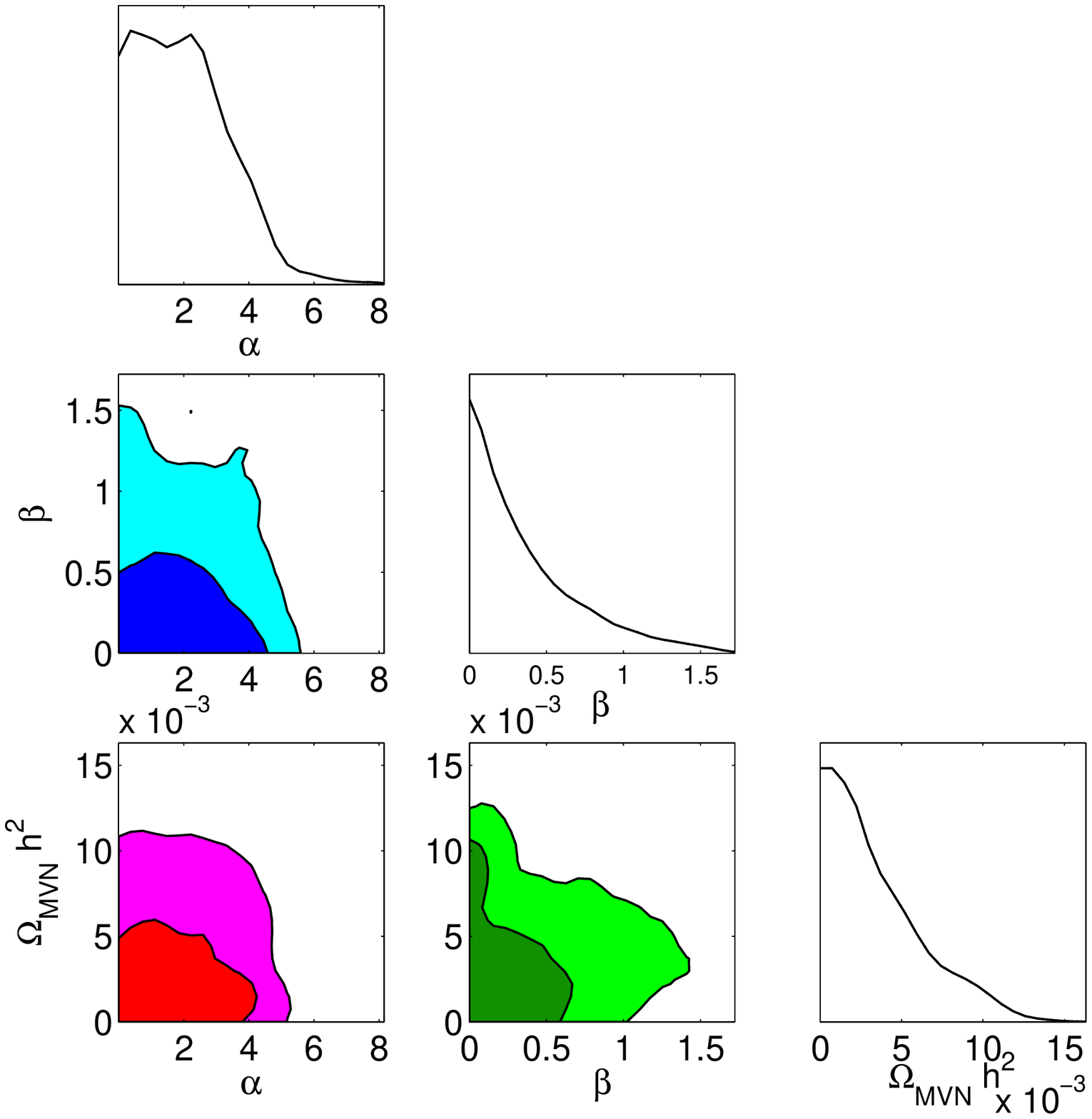}}
    {\includegraphics[width=0.48\textwidth]{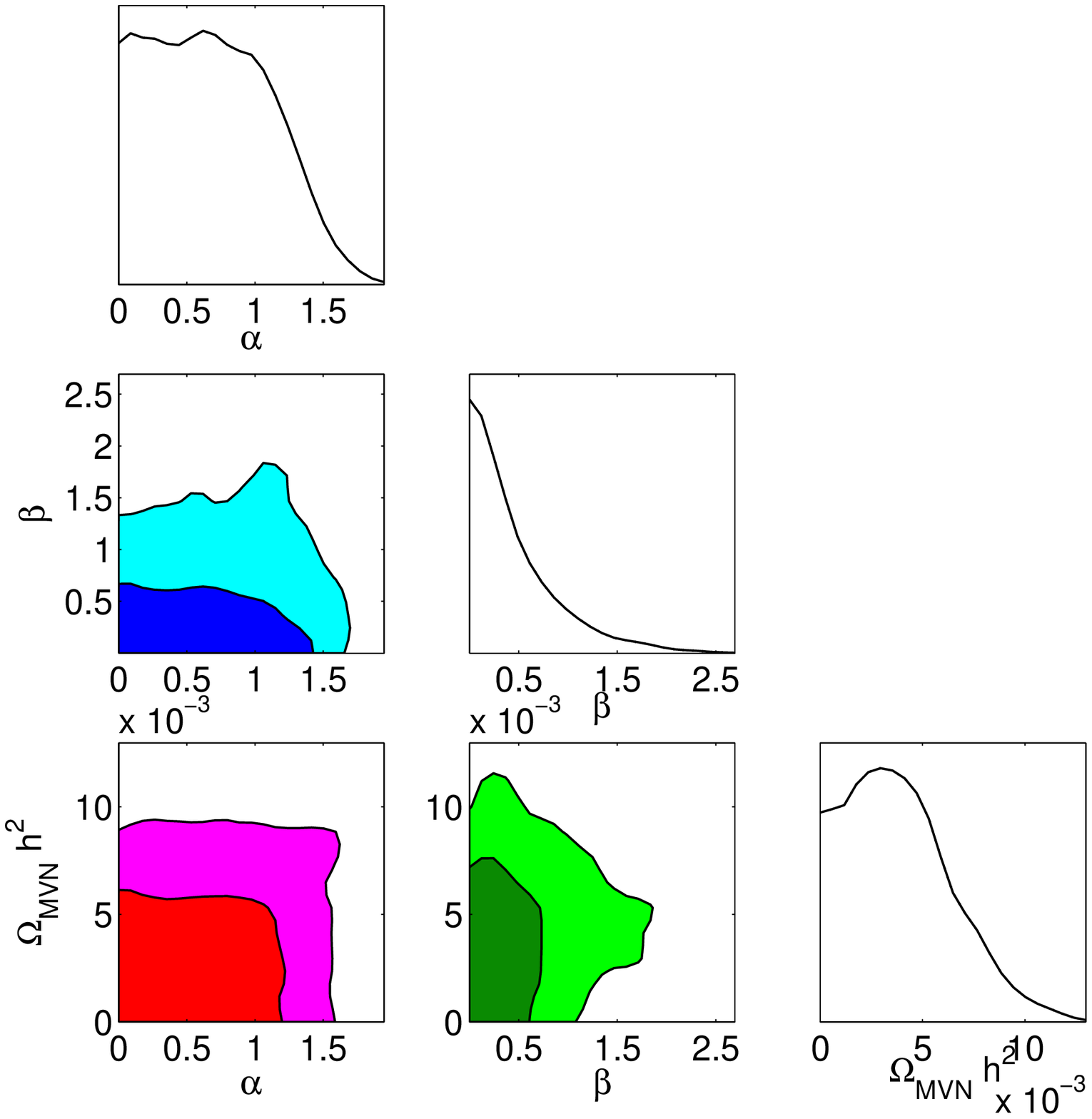}} }
\epsfxsize=5cm
\end{center}
\caption{Contours of constant relative probabilities in two
dimensional
 parameter planes for inverse power law models. Lines correspond to 68\% and 95.4\% confidence limits (left-hand side figure).}
\label{fig:ratra_tri}
\caption{Same as Fig.(\ref{fig:ratra_tri}), but for exponential type
models (right-hand side figure). } \label{fig:EXP_tri}
\end{figure}

Our results are shown in Figs.(\ref{fig:ratra_tri}) -
(\ref{fig:EXP_tri}). In these figures we do not observe the strong
degeneracy between the introduced parameters. This is why one can
put tight constraints on MaVaNu parameters from observations. For
both models we consider, larger $\alpha$ leads larger $w$ at
present. Therefore large $\alpha$ is not allowed due to the same
reason that larger $w$ is not allowed from the current observations.

On the other hand, larger $\beta$ will generally lead larger $m_\nu$
in the early universe. This means that the effect of neutrinos on
the density fluctuation of matter becomes larger leading to the
larger damping of the power at small scales. A complication arise
because the mass of neutrinos at the transition from the
ultra-relativistic regime to the non-relativistic one is not a
monotonic function of $\beta$ as shown in
Fig.(\ref{fig:mass_evolution}). Even so, the coupled neutrinos give
larger decrement of small scale power, and therefore one can limit
the coupling parameter from the large scale structure data.

One may wonder why we can get such a tight constraint on $\beta$,
because it is naively expected that large $\beta$ value should be
allowed if $\Omega_\nu h^2 \sim 0$. In fact, a goodness of fit is
still satisfactory with large $\beta$ value when $\Omega_\nu h^2
\sim 0$, as shown in Fig.({\ref{fig:mass_evolution}). However, the
parameters which give us the best goodness of fit does not mean the
most likely parameters in general. In our parametrization, the
accepted total volume by MCMC in the parameter space where
$\Omega_\nu h^2 \sim 0$ and $\beta\ga 1$ was small, meaning that the
probability of such a parameter set is low.

We find no observational signature which favors the coupling between
MaVaNu and quintessence scalar field, and obtain the upper limit on
the coupling parameter within $2\sigma$ ranges as
\begin{equation}
\beta < 1.11,~ 1.36,~ 1.53~,
\end{equation}
and the present mass of neutrinos is also limited to
\begin{equation}
\Omega_\nu h^2_{\rm{today}} < 0.0095,~ 0.0090,~ 0.0084~,
\end{equation}
for models I, II and III, respectively. 
When we apply the relation
between the total sum of the neutrino masses $M_{\nu}$ and their
contributions to the energy density of the universe:
$\Omega_{\nu}h^2=M_{\nu}/(93.14 eV)$, we obtain the constraint on
the total neutrino mass: $M_{\nu} < 0.87 eV (95 \% C.L.)$ in the
neutrino probe dark-energy model. The total neutrino mass
contributions in the power spectrum is shown in Fig
\ref{fig:nu-mass-PS}, where we can see the significant deviation
from observation data in the case of  large neutrino masses.

\begin{table}[t]
\tbl{Global analysis data within $1\sigma$ deviation for different
types of the quintessence potential.}
{\begin{tabular}{@{}c|c|c|c|c@{}} \toprule
Quantites & Model I
&Model II & Model III & WMAP-3 data ($\Lambda$CDM)  \\ \colrule
$\Omega_B\, h^2[10^2]$ & $2.21\pm 0.07$ & $2.22\pm 0.07$ & $2.21\pm 0.07$ & $2.23\pm 0.07$   \\
$\Omega_{CDM}\, h^2[10^2]$ & $11.10\pm 0.62$ & $11.10\pm 0.65$ & $11.10 \pm 0.63$ & $12.8\pm 0.8$ \\
$H_0$ & $65.97 \pm 3.61$ & $65.37\pm 3.41$ & $65.61\pm 3.26$ & $72\pm 8$  \\
$Z_{re}$ & $10.87 \pm 2.58$ & $10.89 \pm 2.62$ & $11.07 \pm 2.44$ &   ---     \\
$\alpha$ & $< 2.63$ & $< 7.78$ & $< 0.92$ & ---       \\
$\beta$ & $< 0.46 $ & $< 0.47$ & $< 0.58$ & ---       \\
$n_s$ & $0.95\pm 0.02$ & $0.95\pm 0.02$ & $0.95\pm 0.02$ & $0.958\pm 0.016$      \\
$A_s[10^{10}]$ & $20.66\pm 1.31$ & $20.69\pm 1.32$ & $20.72\pm 1.24$ & ---- \\
$\Omega_{Q}[10^2]$ & $68.54\pm 4.81$ & $67.90\pm 4.47$ & $68.22\pm 4.17$ &
$71.6\pm 5.5$      \\
$Age/Gyrs$ & $13.95\pm 0.20$ & $13.97\pm 0.19$ & $13.69\pm 0.19$ &  $13.73\pm 0.16$ \\
$\Omega_{MVN}\,h^2[10^2]$ & $< 0.44$ & $< 0.48$ & $< 0.48$ &  $< 1.97 (95\% C.L.)$ \\
$\tau$ & $0.08\pm 0.03$ & $0.08 \pm 0.03$ & $0.09 \pm 0.03$ & $0.089 \pm 0.030$  \\ \botrule
\end{tabular} \label{ta1}}
\end{table}


\begin{center}
\begin{figure}
\vspace{0cm}
\epsfxsize=5cm
\centerline{
\rotatebox{-90}{\includegraphics[width=0.40\textwidth]{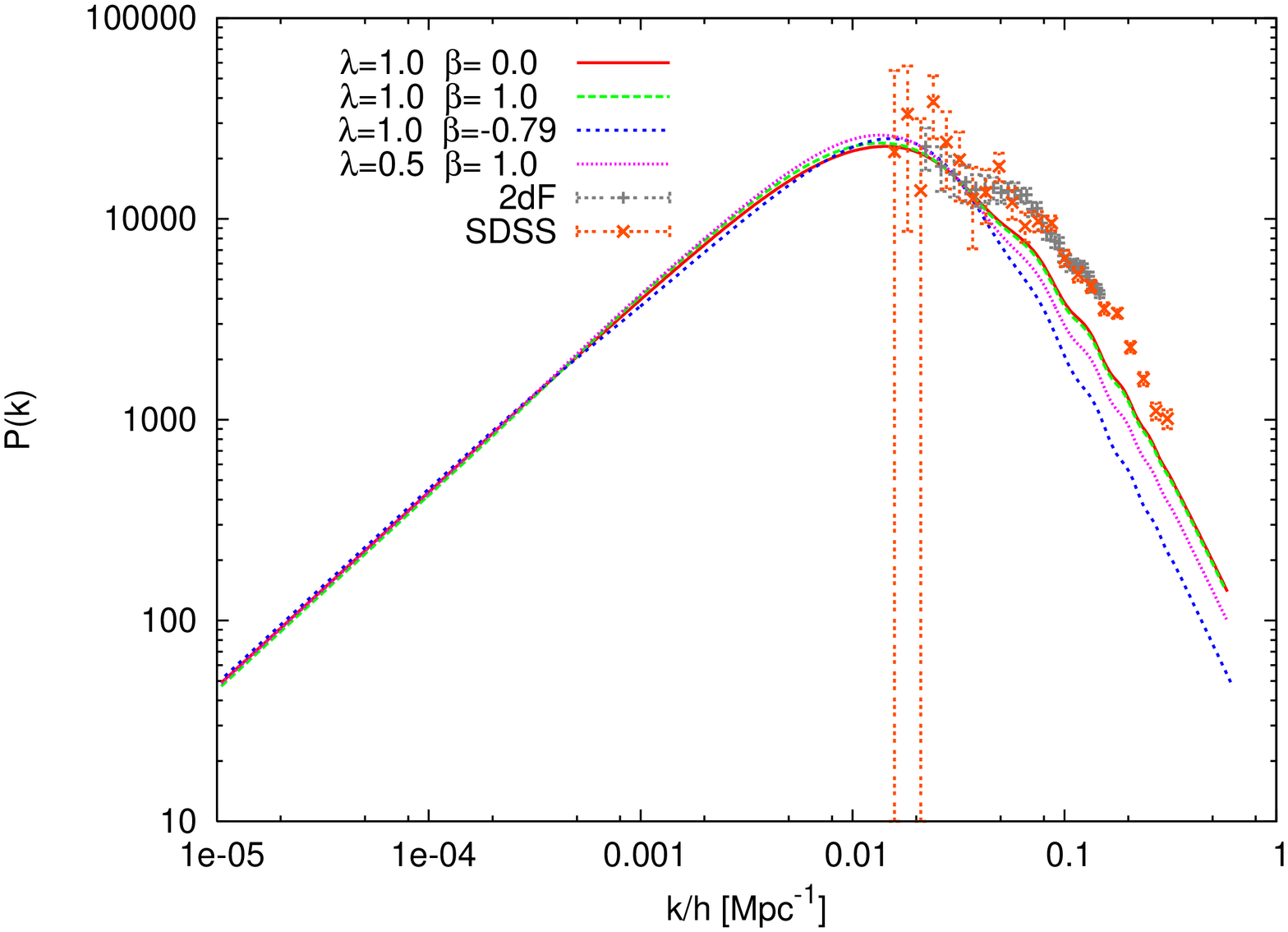}}
\rotatebox{-90}{\includegraphics[width=0.40\textwidth]{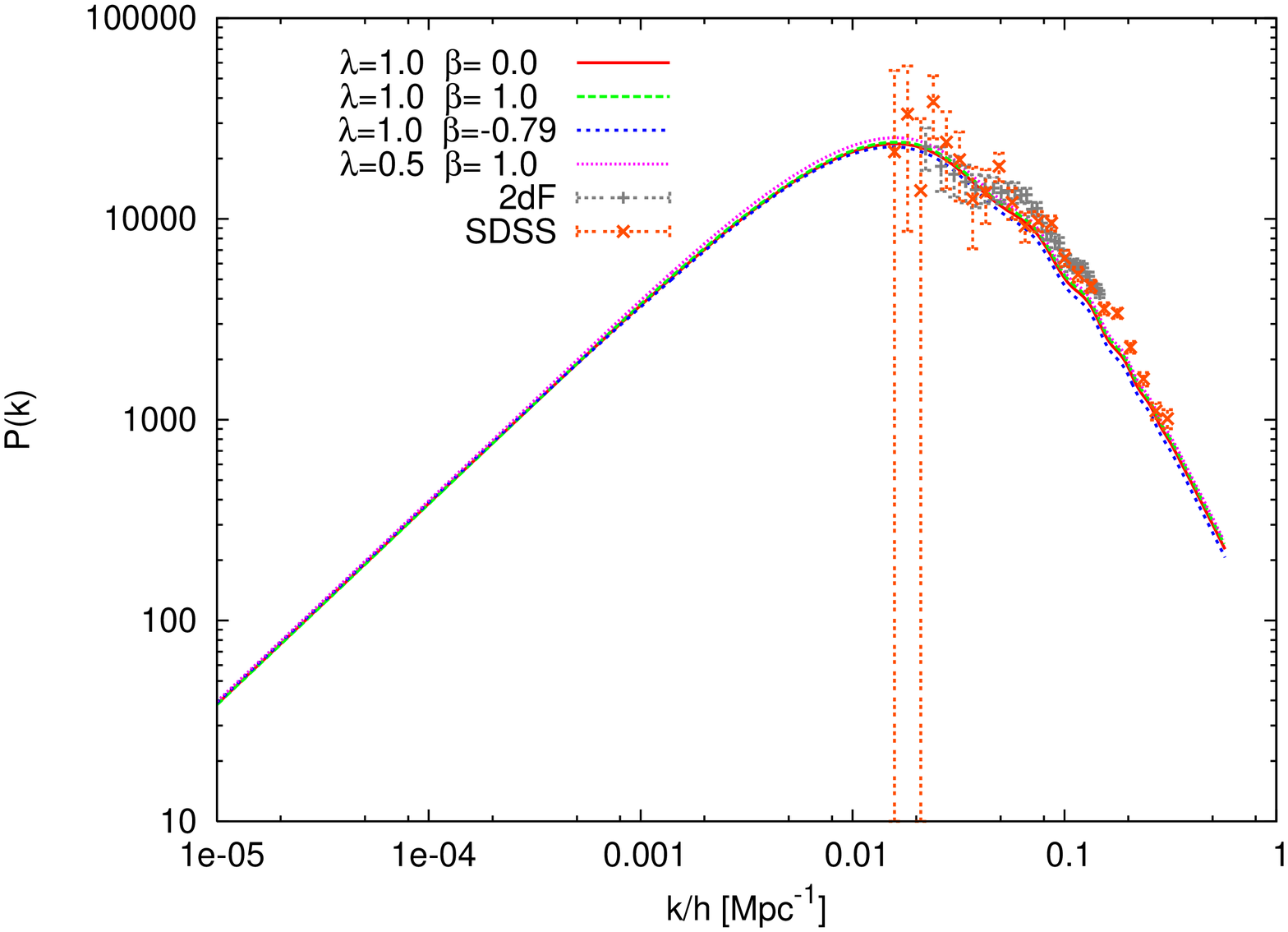}} }
\epsfxsize=5cm
\caption{Examples of the total neutrino mass contributions in power spectrum
 with $M_{\nu}=0.9\, eV$(left-hand side graph) and with $M_{\nu}=0.3\, eV$(right-hand side graph).
Here the variable $\lambda$ is equal to $\alpha$.
 \label{fig:nu-mass-PS} }
\end{figure}
\end{center}
In summary, we investigate dynamics of dark energy in mass-varying
neutrinos. We show and discuss many interesting aspects of the
interacting dark-energy with neutrinos scenario: (1) To explain the
present cosmological observation data, we don't need to tune the
coupling parameters between neutrinos and quintessence field, (2)
Even with a inverse power law potential or exponential type
potential which seem to be ruled out from the observation of
$\omega$ value, we can receive that the apparent value of the
equation of states can pushed down lesser than -1, (3) As a
consequence of global fit, the cosmological neutrino mass bound
beyond $\Lambda CDM$ model was first obtained with the value $\sum
m_{\nu} < 0.87 \, eV (95 \% CL)$.

\section*{Acknowledgments}

We thank to K.~Ichiki for a nice collaboration and many fruitful
discussions. We acknowledge the partial support by CHEP/Kyungbook
National University, Asia Pacific Center for Theoretical
Physics(APCTP) in Korea, and ROC National Science Council in Taiwan.


\begin{thebibliography}{999}
\bibitem{sn1a}
Perlmtter et al.,Nature {\bf 391} (1998) 51[arXiv:astro-ph/7912212];
Riess et al., Astrophys. J. {\bf 116} (1998) 1009[arXiv:astro-ph/980520];
Perlmtter et al., ApJ {\bf 517} (1999) 565[arXiv:astro-ph/9812133].
\bibitem{wmap}
C.~L.~Bennett et al., Astrophys. J. Suppl. Ser. {\bf 148} (2003) 1;
J.~L.~Tonry et al.,Astrophys. J. {\bf 594} (2003) 1;
M.~Tegmark et al., Astrophys. J. {\bf 606} (2004) 702.
\bibitem{lambda}
L.~M.~Krauss and M.~S. Turner, Gen. Rel. Grav. {\bf 27} (1995) 1137 ;
P.~J.~E. Peebles and B.~Ratra, Reviews of Modern Physics, Vol{\bf 75} (2003) 559.
\bibitem{quintessence}
C.~Wetterich, Nucl. Phys. B{\bf 302} (1988) 645;
P.~J.~E. Peebles and B.~Ratra, Astrophys. J. Lett. {\bf 325} (1988) 17.
\bibitem{mgrav}
S.~M.~Carroll, M.~Trodden and M.~S.~Turner, Phys. Rev. D{\bf 70}:
043528 (2004).
\bibitem{mavanu}
D.~B.~Kaplan, A.~E.~Nelson and N.~Weiner, Phys. Rev. Lett. {\bf
93}:091801, (2004); R.~D.~Peccei, Phys. Rev. {\bf D71}:023527
(2005).
\bibitem{review-DE}
E.~J. Copeland, M.~Sami, and S.~Tsujikawa, Int. J. Mod. Phys., {\bf
D15}: 1753 (2006).

\bibitem{k-essence}
C.~A.~Picon, V.~F.Mukhanov, P.~J. Steinhardt, Phys. Rev. {\bf D63}:
103510, 2001
\bibitem{phantom}
R.~R.~Caildwell, Phys. Lett. {\bf B 545}: 23, (2002).
\bibitem{quintom}
Z.~K. Guo,Y.-S. Piao, X.-M. Zhang and Y.-Z Zhang, Phys. Lett. {\bf B 608}, 177 (2005).
\bibitem{seljak:0604335}
U.~Seljak, ~A. Slosar and P.~McDonald, JCAP 0610:014 (2006).

\bibitem{Fardon:2003eh}
R.~Fardon, A.~E.~Nelson and N.~Weiner, JCAP 0410:005, 2004;
[arXiv:astro-ph/0309800].

\bibitem{Bi:2003yr}
X.~J.~Bi, P.~h.~Gu, X.~l.~Wang and X.~M.~Zhang, Phys. Rev. {\bf
D69}:113007 (2004);
[arXiv:hep-ph/0311022].
\bibitem{ostriker03}
J.~P.~Ostriker and P.~Steinhardt,
Science, {\bf 300}, 1909 (2003).

\bibitem{kiyo-keum}
K.~Ichiki and Y.-Y. Keum,
"Primordial Neutrinos, Cosmological Perturbations in Interacting
Dark-Energy Models: CMB and LSS"[arXiv:astro-ph/07052134];
"Cosmological Bounds on Dark Energy-Neutrino Ineractions" (preparing draft).

\bibitem{Anderson:1997un}
  G.~W.~Anderson and S.~M.~Carroll,
  arXiv:astro-ph/9711288.

\bibitem{Brookfield-b}
A.~W.~Brookfield, C.~van de Bruck, D.~F.~Mota, and D.~Tocchini-Valentini,
Phys. Rev. Lett , {\bf 96}: 061301,2006;
Phys. Rev. {\bf D73}:083515,2006.

\bibitem{Spergel:2003cb}
D.~N.~Spergel {\it et al.},
Astrophys. J. Suppl., {\bf 148}, 175 (2003).

\bibitem{Hinshaw:2003ex}
G.~Hinshaw {\it et al.},
Astrophys. J. Suppl.,  {\bf 148}, 135 (2003).

\bibitem{Kogut:2003et}
A.~Kogut {\it et al.},
Astrophys. J. Suppl.,  {\bf 148}, 161 (2003).

\bibitem{Verde:2003ey}
L.~Verde {\it et al.},
Astrophys. J. Suppl.,  {\bf 148}, 195 (2003).

\bibitem{Ma:1995ey}
C.~P.~Ma and E.~Bertschinger,
Astrophys.\ J.\  {\bf 455}, 7 (1995).


\bibitem{camb}
A.~Lewis, A.~Challinor, and A.~Lasenby,
Astrophys. J., {\bf 538}, 473 (2000).

\bibitem{cmbfast}
U.~Seljak and M.~Zaldarriago,
Astrophys. J., {\bf 469}, 437 (1996).

\bibitem{hu}
W.~Hu, D.~Scott, N.~Sugiyama, and M.~White,
Phys. Rev. {\bf D52}, 5498 (1995).

\bibitem{Hinshaw:2006ia}
  G.~Hinshaw {\it et al.}(WMAP collaboration),
  arXiv:astro-ph/0603451.

\bibitem{Page:2006hz}
  L.~Page {\it et al.}(WMAP collaboration),
  arXiv:astro-ph/0603450.

\bibitem{Cole:2005sx}
  S.~Cole {\it et al.}  [The 2dFGRS Collaboration],
  Mon.\ Not.\ Roy.\ Astron.\ Soc.\  {\bf 362}, 505 (2005)
  [arXiv:astro-ph/0501174].

\bibitem{MCMC}
A.~Lewis and S.~Bridle,
 Phys. Rev. {\bf D66}, 103511 (2002).

\bibitem{Ratra:1987rm}
  B.~Ratra and P.~J.~E.~Peebles,
  Phys.\ Rev.\ D {\bf 37}, 3406 (1988).

\bibitem{Brax:1999gp}
  P.~Brax and J.~Martin,
  Phys.\ Lett.\ B {\bf 468}, 40 (1999)
  [arXiv:astro-ph/9905040].


\bibitem{Brax:2000yb}
  P.~Brax, J.~Martin and A.~Riazuelo,
  Phys.\ Rev.\ D {\bf 62}, 103505 (2000)
  [arXiv:astro-ph/0005428].

\bibitem{Copeland:1997et}
  E.~J.~Copeland, A.~R.~Liddle and D.~Wands,
  Phys.\ Rev.\ D {\bf 57}, 4686 (1998)
  [arXiv:gr-qc/9711068].

\bibitem{McDonald:1999dt}
  P.~McDonald, J.~Miralda-Escude, M.~Rauch, W.~L.~W.~Sargent, T.~A.~Barlow, R.~Cen and J.~P.~Ostriker,
  Astrophys.\ J.\  {\bf 543}, 1 (2000)
  [arXiv:astro-ph/9911196].

\bibitem{Croft:2000hs}
  R.~A.~C.~Croft {\it et al.},
  Astrophys.\ J.\  {\bf 581}, 20 (2002)
  [arXiv:astro-ph/0012324].


\bibitem{Goobar:2006xz}
  A.~Goobar, S.~Hannestad, E.~Mortsell and H.~Tu,
  JCAP {\bf 0606}, 019 (2006)
  [arXiv:astro-ph/0602155].

\bibitem{McDonald:2004xp}
  P.~McDonald, U.~Seljak, R.~Cen, P.~Bode and J.~P.~Ostriker,
  Mon.\ Not.\ Roy.\ Astron.\ Soc.\  {\bf 360}, 1471 (2005)
  [arXiv:astro-ph/0407378].
\end{thebibliography}
\end{document}